\documentclass[sigconf]{acmart}

\usepackage{graphicx}
\usepackage{amsmath}
\usepackage{booktabs}
\usepackage{algorithm}
\usepackage{algorithmic}
\usepackage{amsfonts}
\usepackage{multirow}
\usepackage{makecell}
\usepackage{subfigure}
\usepackage{color}
\usepackage{bm}
\usepackage{epstopdf}
\usepackage{url}
\usepackage[cal=cm]{mathalfa}
\usepackage{balance}
\usepackage{threeparttable}
\usepackage{lipsum}
\usepackage{enumitem}

\AtBeginDocument{%
  \providecommand\BibTeX{{%
    \normalfont B\kern-0.5em{\scshape i\kern-0.25em b}\kern-0.8em\TeX}}}

\author{Wentao Xu$^\ast$}
\affiliation{
  \institution{Nanjing University of Science and Technology}
  \country{xuwentao@njust.edu.cn}
 }

\author{Qianqian Xie$^\ast$}
\affiliation{
  \institution{Nanjing University of Science and Technology}
  \country{xieqianqian@njust.edu.cn}
 }

\author{Shuo Yang$^\ast$}
\affiliation{
  \institution{Nanjing University of Science and Technology}
  \country{yangshuo11@njust.edu.cn}
 }

\author{Jiangxia Cao$^\star$$^\S$}
\thanks{$^\ast$Equal contributions to this work}
\thanks{$^\star$Corresponding authors.}
\thanks{$^\S$The publication does not relate to the author’s work at Kuaishou.}
\affiliation{
  \institution{Kuaishou Technology}
  \country{caojiangxia@kuaishou.com}
 }

\author{Shuchao Pang$^\star$}
\affiliation{
  \institution{Nanjing University of Science and Technology}
  \country{pangshuchao@njust.edu.cn}
 }

\copyrightyear{2024} 
\acmYear{2024} 
\setcopyright{acmlicensed}\acmConference[CIKM '24]{Proceedings of the 33rd
ACM International Conference on Information and Knowledge
Management}{October 21--25, 2024}{Boise, ID, USA}
\acmBooktitle{Proceedings of the 33rd ACM International Conference on
Information and Knowledge Management (CIKM '24), October 21--25, 2024,
Boise, ID, USA}
\acmDOI{10.1145/3627673.3679913}
\acmISBN{979-8-4007-0436-9/24/10}

\begin{document}
\title{Enhancing Content-based Recommendation via\\Large Language Model}

\renewcommand{\shorttitle}{LoID}

\begin{abstract}
In real-world applications, users express different behaviors when they interact with different items, including implicit click/like interactions, and explicit comments/reviews interactions.
%
Nevertheless, almost all recommender works are focused on how to describe user preferences by the implicit click/like interactions, to find the synergy of people.
For the content-based explicit comments/reviews interactions, some works attempt to utilize them to mine the semantic knowledge to enhance recommender models. However, they still neglect the following two points:
(1) The content semantic is a universal world knowledge; how do we extract the multi-aspect semantic information to empower different domains?
(2) The user/item ID feature is a fundamental element for recommender models; how do we align the ID and content semantic feature space?
%
%
In this paper, we propose a `plugin' semantic knowledge transferring method \textbf{LoID}, which includes two major components: (1) LoRA-based large language model pretraining to extract multi-aspect semantic information; (2) ID-based contrastive objective to align their feature spaces. 
%
We conduct extensive experiments with SOTA baselines to demonstrate superiority of our method LoID.
\end{abstract}



\maketitle

\section{Introduction}

\textbf{Background.}
With the boom of digital information, billions of user requests are produced daily, so recommender systems (RSs) have become an integral part of Internet platforms.
To capture users' interests more accurately, RSs have gone through several milestones, such as the logistic regression with hand-crafted features (e.g., FM~\cite{fm}), the neural networks (e.g., WideDeep~\cite{widedeep}, YoutubeNet~\cite{youtubenet}), the sequential signal (e.g., DIN~\cite{din}, SIM~\cite{sim}), and the multi-hop graph signal (e.g., PinSage~\cite{pinsage}, DGRec~\cite{DGRec}).
In retrospect, these effective methods are based on the collaborative filtering (CF) idea and extend the boundaries of RSs. However, the CF framework also limits them under the case of cold-start and data sparsity problems.
The reason is that the CF idea aims to mine the user/item pattern intelligence from data, to discover and recommend high-click candidate items for a user while it is hard to understand the users' fine-grained and multi-aspect interests. 
In fact, instead of mining user preferences from massive user-item interaction logs, the user always leaves some reviews/comments to explain further his/her feelings about this interaction, which provides an explicit way to understand the users' complex interests in language semantic space.
And the recent effort MoRec~\cite{id} claimed that multi-modal information yield better results in sequential recommendation.

\textbf{Related work.} 
%
To extract valuable content semantic information, the pioneering works are formed as a rating prediction task: for a user-item pair in test set, give the historical user/item contents in training set (e.g., reviews), then predict their possible interaction rating.
In early years, the DeepCoNN~\cite{deepconn} employed two convolutional neural networks (CNNs) towers to aggregate user/item content tokens individually to measure their dot-similarity, and the D-Attn~\cite{dattn} further extended DeepCoNN by introducing local and global attention mechanism to replace CNNs to aggregate tokens.
Following the D-Attn, the ALFM~\cite{alfm} and ANR~\cite{anr} focused on extracting the fine-grained multi-aspect semantic information and assigning different weights for aspects.
The recent progress is RGCL~\cite{RGCL}, which used BERT to generate the user-item content scores and then leveraged them as user-item graph edge weights to conduct a multi-hop graph neural network to make prediction.
%
%
%
%

\textbf{Motivation.}
Although these methods raised model ability with content information, they ignore the following problems:
\begin{itemize}[leftmargin=*,align=left]
    \item \textit{Transfer semantic knowledge across domains}: Actually, RS needs to serve several domains simultaneously, such as electronics, clothing, books, etc.
    Since different domains always express different aspects of users' interests (e.g., food is delicious, price is appropriate, etc.), the previous methods need to re-tune their semantic component for different domains, which is time-consuming.
    \item \textit{Enhancing the correlation between content and ID}: 
    The user-item content and ID-interaction information can be seen as two different modalities~\cite{qformer} connected by users.
    Nevertheless, previous methods focus on utilizing separate components to model the two corresponding content/ID spaces while ignoring how to exploit the correlation and align them in a unified space.
    %
\end{itemize}

\textbf{Our Work.}
To alleviate the above problems, we propose \textbf{LoID}, a LLM-based~\cite{gpt-3} model for transferring semantic knowledge across domains based on LoRA~\cite{lora}, and aligning content/ID information with contrastive objectives~\cite{clip}. 
It mainly includes two steps:
\begin{enumerate}[leftmargin=*,align=left]
\item \textit{Pretraining semantic `plugin'}: On the one hand, the ideal semantic information should act as a universal role to support all domains. On the other hand, different domains have their main aspects of semantic information that are related to recommendation. 
We borrow the LoRA strategy idea to train a small set of parameters for each domain, which could serve as plugins. 
Then we fused them by DARE ~\cite{DARE}without further re-training.

\item \textit{Aligning the content/ID feature spaces}: As discussed before, the content and ID information can be seen as two modalities of user-item interaction.
To minimize their gap, we introduce the contrastive idea of maximizing content/ID features' mutual information to align their feature spaces.

\end{enumerate}
%
%
Finally, to validate our LoID effectiveness, we extensively test LoID under 11 different domain datasets to show its superior ability.

\textbf{Contributions.} 
Our contributions are summarized as follows:
\begin{itemize}[leftmargin=*,align=left]
\item We give a `plugin' idea to transfer semantic knowledge, which sheds light on building a new paradigm for recommendation\footnote{Codes at \url{https://github.com/cjx96/LoID}.}.
\item We devise a novel content/ID feature alignment objective.
\item We conducted detailed analyses with SOTA methods and LLMs.

\end{itemize}

\section{Preliminary}

\subsection{Problem Statement}
This work considers a brief task: For one domain dataset $\mathcal{D}$, and each user-item interaction primarily comprises the following four elements: user ${\mathrm{u}}$, item ${\mathrm{i}}$, the corresponding rating $r_{\mathrm{u,i}}$, and the textual content token list $\textbf{t}_{\mathrm{u,i}}$ left by the user. 
Our model aims to predict ratings by historical content and user/item ID information. 

Besides, \textbf{to test the effectiveness of the  `plugin' idea , we further consider multi-domain scenarios}, that is: given source domain content information $\{(\textbf{t}^{\mathrm{source_1}}, r^{\mathrm{source_1}}), (\textbf{t}^{\mathrm{source_2}}, r^{\mathrm{source_2}}), \dots \}$, predict target domain $\mathcal{D}_{\mathrm{target}}$ rating score.

\subsection{Low-Rank Adaptation (LoRA)}
Before going on, we first explain the LoRA~\cite{lora}, to show the basic idea of how to fast tuning an LLM~\cite{bert}.
Indeed, the unit block of LLM, transformer~\cite{attention}, consists of two parts: (Masked) attention and MLPs (FFN).
The two parts introduce four pre-trained parameter matrices (e.g., $\mathbf{W}_{\mathrm{Q}}, \mathbf{W}_{\mathrm{K}}, \mathbf{W}_{\mathrm{V}}, \mathbf{W}_{\mathrm{FFN}}$).
Then, the challenge is: how to update those matrices without re-training them or involving new large parameters.
Thereby, LoRA was proposed by assigning two small matrices for any pre-trained matrix $\mathbf{W}_{\mathrm{LLM}}^{\mathrm{Frozen}} \in \theta ^ \mathrm{Frozen}_\mathrm{LLM}$:
%
%
\begin{equation}
\footnotesize
 \begin{split}
\mathbf{W}^{\mathrm{LoRA}} = \mathbf{W}_{\mathrm{LLM}}^{\mathrm{Frozen}} + \mathbf{B}^{\mathrm{LoRA}} \mathbf{A}^{\mathrm{LoRA}},
 \end{split}
\end{equation}
where $\mathbf{W}^{\mathrm{LoRA}}\in \mathbb{R}^{d \times d}$ is the modified weight matrix,
$\mathbf{B}^{\mathrm{LoRA}}\in \mathbb{R}^{d \times r}$ and $\mathbf{A}^{\mathrm{LoRA}}\in \mathbb{R}^{r \times d}$ are low-rank matrices$(r<<d)$.
LoRA freezes the original large parameter matrix $\mathbf{W}$, introduces and updates it by $\mathbf{B}^{\mathrm{LoRA}}\mathbf{A}^{\mathrm{LoRA}} \in \mathbb{R}^{d \times d}$.
%
Next, we can fine-tune the LLM by any supervised $\textit{Task}$ that without updating all parameters as follows:
\begin{equation}
\footnotesize
\begin{split}
\theta^\mathrm{LoRA}_\mathrm{Task} = \mathrm{LoRA(}\theta ^\mathrm{Frozen}_\mathrm{LLM},\mathrm{Task}), \quad \theta_\mathrm{Task}=\theta^\mathrm{Frozen}_\mathrm{LLM}+\theta^\mathrm{LoRA}_\mathrm{Task},
 \end{split}
\label{lora}
\end{equation}

where $\theta^\mathrm{LoRA}_\mathrm{Task}$ is extra added parameters and $\theta_\mathrm{Task}$ is all parameters\footnote{In our work, we only introduce extra 4M $\theta^\mathrm{LoRA}_\mathrm{Task}$ parameters to tune 110M $\theta ^\mathrm{pre}_\mathrm{BERT}$ BERT.}.
%
Among different domains, $\theta^\mathrm{LoRA}_\mathrm{Task}$ has the same structure, so it is a proper semantic component for knowledge transfer ~\cite{adapter}.

\subsection{Drop and Rescale (DARE)}
Assuming that we have fine-tuned the same LLM in different specific tasks and obtained several (supervised fine-tuned) SFT models as $\{\theta_\mathrm{Task_1}, \theta_\mathrm{Task_2}, \dots\}$, a challenge is: how to merge them as an omnipotent LLM?
To answer it, DARE~\cite{DARE} was proposed, which devises a drop-then-rescale strategy on the gradient update sets $\{\delta_{\mathrm{Task_1}}, \delta_{\mathrm{Task_2}}, \cdots\}$ to make the multiple homogenous LLM merging, here we take the LoRA-Style SFT as an example:
\begin{equation}
\footnotesize
\begin{split}
    \delta_{\mathrm{Task_i}}=\theta_\mathrm{Task_i}-\theta_\mathrm{LLM} = \mathbf{B}^{\mathrm{LoRA}}_{\mathrm{Task_i}} \mathbf{A}^{\mathrm{LoRA}}_{\mathrm{Task_i}}\\
 \end{split}
\label{delta parameter}
\end{equation}

%
%
Next, DARE performs random drop on $\delta_{\mathrm{Task}}$ with probability $p$, and then rescales the remaining ones by a factor $1/(1 - p)$.
\begin{equation}
\footnotesize
\begin{split}
\tilde{\delta}_\mathrm{Task_{i}}=\mathrm{Drop}(\delta_{\mathrm{Task_{i}}}), \quad 
\tilde{\delta}^\mathrm{DARE}_\mathrm{Task_{i}}=\tilde{\delta}_\mathrm{Task_{i}}/(1-p)
 \end{split}
\label{dare}
\end{equation}
where $p$ can reach 0.9 even 0.99,  $\tilde{\delta}_\mathrm{Task_{i}}$ and $\tilde{\delta}^\mathrm{DARE}_\mathrm{Task_{i}}$ represent the delta parameters after dropping and rescaling respectively. 
 \begin{equation}
 \footnotesize
 \begin{split}
 \theta^\mathrm{DARE}_\mathrm{LLM} &= \mathrm{DARE(}\theta ^\mathrm{Frozen}_\mathrm{LLM}, \theta_\mathrm{Task_{1}},\theta_\mathrm{Task_{2}},\cdots,\theta_\mathrm{Task_{n}}) \\
 &= \theta_\mathrm{LLM}^\mathrm{Frozen} + \tilde{\delta}^\mathrm{DARE}_\mathrm{Task_{1}} + \tilde{\delta}^\mathrm{DARE}_\mathrm{Task_{2}} + \tilde{\delta}^\mathrm{DARE}_\mathrm{Task_{3}} + \cdots 
\end{split}
\label{multi-lora}
\end{equation}
where $\theta^{\mathrm{DARE}}_\mathrm{LLM}$ is the merged SFT LLM.

\section{Methodology}

\subsection{Overview}
\begin{figure}[t]
\begin{center}
\includegraphics[width=8cm,height=6cm]{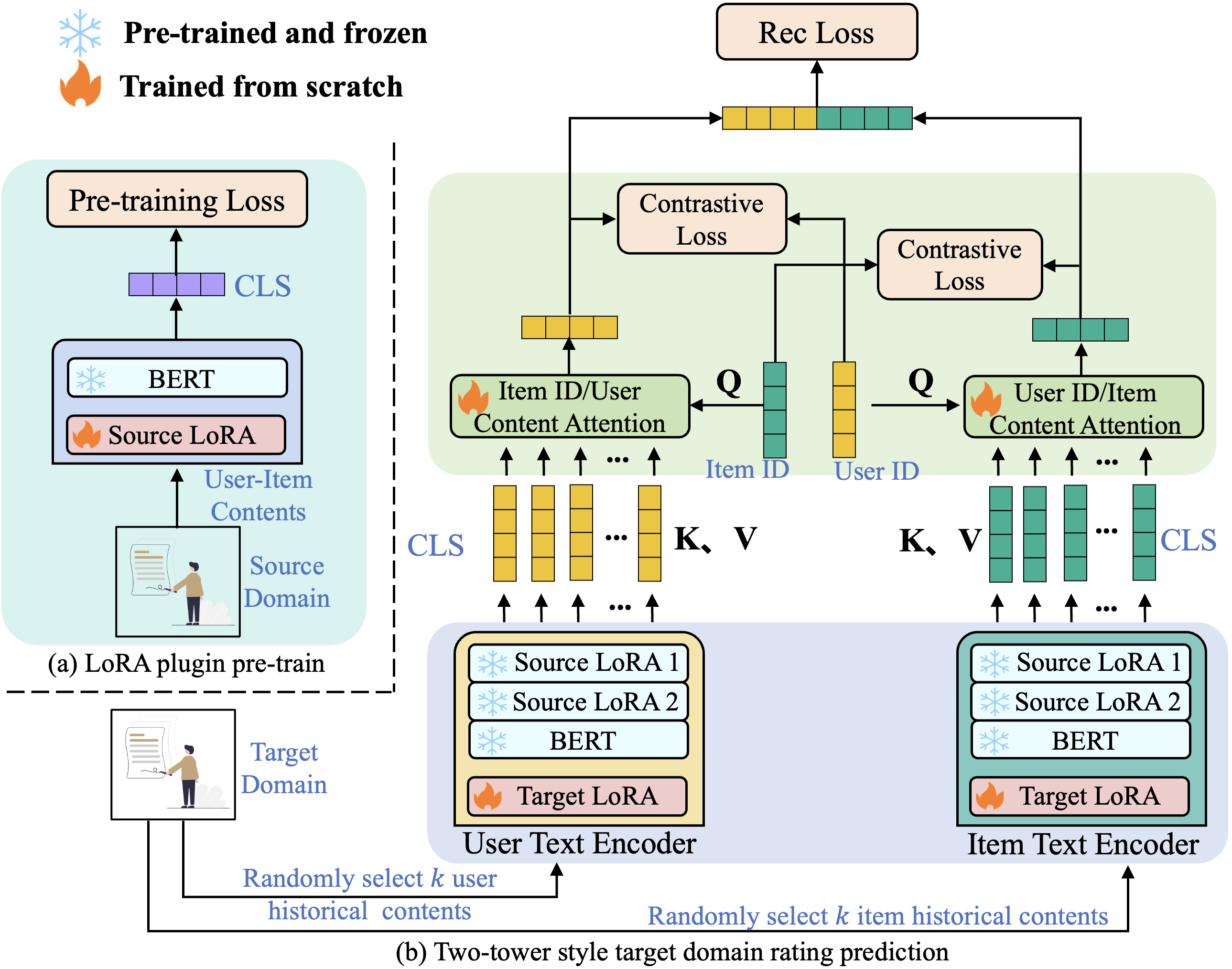}
\caption{Model Architecture of the proposed LoID.}
\label{ourmethod}
\end{center}
\vspace{-0.5cm}
\end{figure}

The architecture of our method LoID is illustrated in Figure~\ref{ourmethod}.
%
In part (a), we train the LoRA parameters of source domains as "plug-ins" to enhance target domain prediction without further re-training. 
In part (b), we extract historical contents of user/item to obtain user/item semantic representation, and then align the ID and semantic to make target domain rating prediction, \textbf{note that the source domain LoRA plugin is an optional choice}. 

\subsection{Source Domain LoRA Pretraining}


\textbf{Source LoRA.}
To get users' behavior data from the source domain, We leverage the LoRA strategy to pre-train the source domain.
%
In the pre-training task, we leverage rating prediction by $\mathbf{CLS}_\mathrm{source}$ token to predict rating directly, here we take BERT as default LLM:

\begin{equation*}
 \footnotesize
 \begin{split}
 \theta^\mathrm{LoRA}_\mathrm{source_{k}} = \mathrm{LoRA(}\theta ^\mathrm{Frozen}_\mathrm{BERT},\mathrm{Task}_{\mathrm{source_{k}}})&, \ \ \ 
  \theta_\mathrm{source_{k}}=\theta ^\mathrm{Frozen}_\mathrm{BERT}+\theta^\mathrm{LoRA}_\mathrm{source_{k}}, \\
 \mathbf{CLS}^\mathrm{source}_{\mathrm{u,i}} = \text{BERT}(\textbf{t}_{\mathrm{u,i}}^\mathrm{source_{k}}, \theta_\mathrm{source_{k}}),& \quad 
\hat{r}_{\mathrm{u,i}}^\mathrm{source_{k}} = \text{Predict}(\mathbf{CLS}^\mathrm{source_{k}}_{\mathrm{u,i}}), \\
\end{split}
\label{predict}
\end{equation*}
where $k$ denotes the $k^{th}$ domain in the source domains.
%
where \text{BERT}$(\cdot)$ is the LLM forward procedure, \text{Predict}$(\cdot)$ is an MLP to generate the prediction score $\hat{r}_{\mathrm{u,i}}^\mathrm{source}$, then we adopt the Mean Squared Error (MSE) loss to optimize LoRA parameters:
 \begin{equation}
\footnotesize
\begin{split}
\mathcal{L}_\mathrm{pre-train}^\mathrm{source}= \frac{1}{|\mathcal{D}_{\mathrm{source}}|} \sum_\mathrm{(u, i)}^{\mathcal{D}_{\mathrm{source}}} (\hat{r}_\mathrm{u,i}^\mathrm{source} - r_\mathrm{u,i}^\mathrm{source})^2,
\label{L_pre}
\end{split}
\end{equation}

where $|\mathcal{D}_{\mathrm{source}}|$ means the number of the samples of source domains, $\hat{r}_\mathrm{u,i}$/$r_\mathrm{u,i}$ represent predicted/real rating respectively.
%

%

\subsection{Target Domain Re-LoRA}
After obtaining several source semantic `plugin' LoRAs, we then transfer their general modality knowledge to target in two steps.

\textbf{Source LoRA Merging.} 
%
The first step is to merge these pre-trained source plugins with LLM, as discussed before, we follow DARE~\cite{DARE} idea to merge multiple LoRAs parameters.
 \begin{equation}
 \footnotesize
 \begin{split}
 \theta^\mathrm{DARE}_\mathrm{BERT} = \mathrm{DARE(}\theta ^\mathrm{Frozen}_\mathrm{BERT}, \theta_\mathrm{source_{1}},\theta_\mathrm{source_{2}},\cdots,\theta_\mathrm{source_{n}})
\end{split}
\label{multi-lora}
\end{equation}
where $n \ge 0$ represents the avaliable source domains, \textbf{note that the $n=0$ means ignoring all source domain LoRAs}.

\textbf{Target LoRA.} 
Enabling the model to adapt to the target domain, we introduce a LoRA module (i.e., Target LoRA) again in the target domain.
In the Re-LoRA, only the Target LoRA tuning its parameters; the merged BERT parameters are frozen:
\begin{equation*}
 \footnotesize
 \begin{split}
 \theta^\mathrm{LoRA}_\mathrm{target} = \mathrm{LoRA(}\theta ^\mathrm{DARE,Frozen}_\mathrm{BERT},\mathrm{Task}_{\mathrm{target}})&, \ \ \ 
  \theta_\mathrm{target}=\theta ^\mathrm{DARE,Frozen}_\mathrm{BERT}+\theta^\mathrm{LoRA}_\mathrm{target}, \\
\end{split}
\label{lora}
\end{equation*}
where $\theta_\mathrm{target}$ is the LLM part's parameters of target domain.
%

\subsection{ID-based Contrastive Learning }
\label{historical}
As the fundamental element of CF-based recommender work, the user/item IDs are indispensable in achieving personalized signals. 
How to align the independent feature space of semantic/ID is the key to making the content information more competitive for industrial RS.
Therefore, we design an auxiliary task. 

\textbf{User/Item Text Encoder.} Considering the computation scale of LLM, we randomly extract historical ${\mathrm{k}}$ contents in the target domain ($\{\mathbf{t}_\mathrm{u}^\mathrm{1}, \dots,\mathbf{t}_\mathrm{u}^\mathrm{k}\}$/$\{\mathbf{t}_\mathrm{i}^\mathrm{1}, \dots, \mathbf{t}_\mathrm{i}^\mathrm{k}\}$) to describe the target domain user/item ${\mathrm{u}}/{\mathrm{i}}$ holistic preferences/properties: 
\begin{equation}
\footnotesize
\begin{split}
\mathbf{CLS}_{\mathrm{u}}^{\mathrm{k}} = \text{BERT}(\textbf{t}_{\mathrm{u}}^{\mathrm{k}}, \theta _\mathrm{target}), \quad \mathbf{CLS}_{\mathrm{i}}^{\mathrm{k}} = \text{BERT}(\textbf{t}_{\mathrm{i}}^{\mathrm{k}}, \theta _\mathrm{target}),
\end{split}
\end{equation}
where $\mathbf{CLS}_{\mathrm{u}}^{\mathrm{k}}$ and $\mathbf{CLS}_{\mathrm{i}}^{\mathrm{k}}$ are the embeddings for the user's and item's $k^{th}$ content.$\textbf{t}_{\mathrm{u}}^{\mathrm{k}}$ and $\textbf{t}_{\mathrm{i}}^{\mathrm{k}}$ are the corresponding textual contents.

\textbf{Attention Layer.} 
On top of the user/item historical content $\mathbf{CLS}$ semantic information, we devise a novel attention mechanism to exchange the user/item semantic/ID information.
Specifically, we integrate the semantic information $\{\mathbf{CLS}_{\mathrm{u}}^1, \dots, \mathbf{CLS}_{\mathrm{u}}^k\}$ into the item representation $\mathbf{p}_{i}$, versus versa for user representation $\mathbf{p}_{u}$. 
%
%
\begin{equation}
\footnotesize
\begin{split}
\mathbf{v}_i &= \text{Attention}(\mathbf{p}_i, \{\mathbf{CLS}_{\mathrm{u}}^\mathrm{1}, \dots, \mathbf{CLS}_{\mathrm{u}}^\mathrm{k}\}, \{\mathbf{CLS}_{\mathrm{u}}^\mathrm{1}, \dots, \mathbf{CLS}_{\mathrm{u}}^\mathrm{k}\}), \\
\mathbf{v}_u &= \text{Attention}(\mathbf{p}_u, \{\mathbf{CLS}_{\mathrm{i}}^\mathrm{1}, \dots, \mathbf{CLS}_{\mathrm{i}}^\mathrm{k}\}, \{\mathbf{CLS}_{\mathrm{i}}^\mathrm{1}, \dots, \mathbf{CLS}_{\mathrm{i}}^\mathrm{k}\}), \\
\end{split}
\end{equation}
where $\mathbf{v}_i$ and $\mathbf{v}_u$ are the updated item and user representations.
%

 \textbf{Contrastive Loss.} The Attention mechanism primarily focuses on different parts of the input sequence, but it does not consider the relationship between users and items. So, we conduct contrastive learning, enhancing the similarity between interactive users and items.
 For each updated representation $\mathbf{v}_i$/$\mathbf{v}_u$, treated as an anchor, we pair it with the original user/item feature representation as a positive sample $\mathbf{p}^+_u/\mathbf{p}^+_i$, and another representation in the batch as a negative sample $\mathbf{p}^-_u/\mathbf{p}^-_i$. The goal is to minimize the distance $\mathrm{D}(anchor, positive)$ and maximize $\mathrm{D}(anchor, negative)$.
 %
 %
 \begin{equation*}
\footnotesize
\begin{split}
\mathrm{D} &= ||anchor-sample||^2_2, \\
\mathcal{L}_\mathrm{cl} = \max(0, \Delta + \mathrm{D}(\mathbf{v}_i, \mathbf{p}^+_u)& - \mathrm{D}(\mathbf{v}_i, \mathbf{p}^-_u)) 
+ \max(0, \Delta + \mathrm{D}(\mathbf{v}_u, \mathbf{p}^+_i) - \mathrm{D}(\mathbf{v}_u, \mathbf{p}^-_i)), \\
\end{split}
\label{triplet_loss}
\end{equation*}
where ${\mathbf{v}_i}/\mathbf{v}_u$ is the representation of the updated item/user, ${\mathbf{p}^+_u/\mathbf{p}^+_i}$ represents the  positive sample, and ${\mathbf{p}^-_u/\mathbf{p}^-_i}$ represents the negative sample.
${\Delta}$ is a margin constant that enforces a "safe distance" constraint on correctly classified samples.

\subsection{Model Optimization}
%
%

%
%

Finally, we concatenate $\mathbf{v}_u$ and $\mathbf{v}_i$ to predict the rating:
%
\begin{equation}
\footnotesize
\begin{split}
\hat{r}_\mathrm{u,i} = \text{Predict}(\text{Concat}(\mathbf{v}_u,\mathbf{v}_i)), \ \ 
\mathcal{L}_\mathrm{rec} = \frac{1}{{|\mathcal{D}_{\mathrm{target}}|}} \sum_\mathrm{(u, i)}^{\mathcal{D}_{\mathrm{target}}} (\hat{r}_\mathrm{u,i} - r_\mathrm{u,i})^2,
\end{split}
\label{L_rec}
\end{equation}
where $|\mathcal{D}_{\mathrm{target}}$| means the number of the samples, $\hat{r}_{\mathrm{u,i}}$ and $r_{\mathrm{u,i}}$ represent predicted and real rating respectively.

Throughout the training process, the total loss comprises the rating prediction loss and the contrastive loss:
%
\begin{equation}
\footnotesize
\mathcal{L} =\mathcal{L}_\mathrm{rec} + \lambda\mathcal{L}_\mathrm{cl},
\label{custom_loss}
\end{equation}
where $\mathcal{L}_\mathrm{rec}$ denotes the prediction loss, ${\lambda}$ is the weight assigned to the contrastive loss, and ${\mathcal{L}_\mathrm{cl}}$ represents the contrastive loss.


\section{Experiments}
%
\subsection{Experimental Setup}

\begin{table*}[th]
\footnotesize
\centering
\caption{The experimental results on Amazon datasets (MSE, lower is better.)}
\resizebox{\linewidth}{!}{
\setlength{\tabcolsep}{13pt}{
\begin{tabular}{l|cc|cc|cc|cccc}
\toprule
Datasets & DeepConn & D-attn & ALFM & ANR & BiGI & RGCL & LoID(w/o ID CL) & LoID & LoID(Elec) & LoID(Elec \&Movie) \\
\midrule
Electronics & 1.659 & 1.744 &  1.563 & 1.445 & 1.1433 & 0.9621 & 0.8738 & 0.8666 & $\star$ & $\star$ \\
Movies \& TV & 1.207 & 1.246 & 1.193 & 1.112 & 1.3489 & 0.9239 & 0.8526 & 0.8362 & 0.8167 & $\star$ \\
CDs \& Vinyl & 0.980 & 1.014 &  0.956 & 0.914 &1.1831 & 0.8180 & 0.7001 & 0.6558 & 0.6295 & 0.6170\\
\midrule
Amazon Instant Video & 1.178 & 1.213 & 1.075 & 1.009 & 1.2437 & 0.9357 & 0.5097 & 0.5094 & 0.4891 & 0.4731\\
Baby & 1.442 & 1.507 &  1.359 & 1.258& 1.5166 & 1.1414 & 0.7436 & 0.7275 & 0.6911 & 0.6888\\
Digital Music & 0.749 & 0.775 &  0.725 & 0.688 & 0.7731 & 0.7735 & 0.4664 & 0.4709 & 0.4380 & 0.4195\\
Musical Instruments & 1.160 & 1.224 & 1.072 & 1.034 & 1.3580 & 0.7211 & 0.5878 & 0.5816 & 0.5179 & 0.5322\\
Office Products & 1.569 & 1.650 & 1.474 & 1.337 & 1.6902 & 0.7001 & 0.6422
& 0.6016 & 0.5909 & 0.5621\\
Patio, Lawn \& Garden & 1.622 & 1.696 &  1.510 & 1.403 & 1.7205 & 0.7049 & 0.6833 & 0.6703 & 0.6358 & 0.6281\\
Pet Supplies & 1.565 & 1.628 &  1.485 & 1.377 & 1.6185 & 1.2380 & 0.7436
 &  0.7679 & 0.7304 & 0.7196 \\
Video Games& 1.498  & 1.533 & 1.383 & 1.292 & 1.4450 & 1.0826 & 0.7994 & 0.7881 & 0.7541 & 0.7336
\\
\bottomrule
\end{tabular}
}}
\begin{center}
Remark `$\star$' indicates that this domain is served as the source domain. We report results for LoID and its variants under an average of 5 different random seed result settings.
\end{center}
\label{main_exp}
\vspace{-0.3cm}
\end{table*}

\subsubsection{Datasets}
We selected 11 categories~\footnote{\url{http://snap.stanford.edu/data/amazon/productGraph/categoryFiles/}} of representative datasets from the Amazon dataset~\cite{amazon}.
Among them, the smallest dataset Musical Instruments includes 339,231 users, 83,046 items, 500,176 ratings, and the largest dataset Electronics comprises 4,201,696 users, 476,002 items, 7,824,482 ratings.
Considering the different data-scale, we select three largest datasets (e.g., Electronics, Movies, and CDs) as our source domains, which contain relatively rich records of user-item contents and ratings.
%
%
%
Following \cite{anr,transnets,regression,dattn}, we randomly partitioned our datasets into training, validation, and test sets with an 8:1:1 ratio.
%
%

\subsubsection{Baselines}
We compare LoID with three classes of baselines:
%
(1) Single-aspect methods, such as DeepCoNN~\cite{deepconn} and D-Attn~\cite{dattn}, extract features from the historical semantic information of users and items using CNN and attention mechanisms.
(2) Multi-aspect methods, including ALFM~\cite{alfm} and ANR~\cite{anr}, aim to extract multiple semantic aspects and their respective importance.
(3) GNNs-based methods, such as BiGI~\cite{BiGI} and RGCL~\cite{RGCL}, integrate adjacent node information from collaborative filtering.


\begin{table}[t]
    \centering
    \footnotesize
    \caption{Performance comparison of Domain Correlation.}
    \label{comparison_datasets}
    \setlength\tabcolsep{2pt}{
        \begin{tabular}{lcc|cc|cc}
        \toprule
        \multirow{2}{*}{Datasets} & \multicolumn{2}{c|}{Electronics} &\multicolumn{2}{c|}{Movies} &\multicolumn{2}{c}{CDs}\\ \cline{2-7}&Sim & MSE &Sim & MSE &Sim & MSE \\
        \hline
        Amazon Instant Video & 0.07 & 0.489(+5\%) & 0.28 & 0.473(+9\%) & 0.16 & 0.485(+6\%)\\
        Digital Music & 0.08 & 0.438(+6\%) & 0.15 &  0.436(+7\%) & 0.33 & 0.425(+9\%)\\
        Musical Instruments & 0.22 & 0.518(+11\%) & 0.11 & 0.546(+6\%) & 0.21 & 0.540(+7\%)\\
        \bottomrule
        \end{tabular}
    }
\vspace{-0.5cm}
\end{table}

\subsubsection{Parameter Settings}
The embedding size $d$ ,dropout rate,learning rate and batch size is fixed as 768, 0.5, 1e-5 and 4. 
%
In the LoRA module, the low-rank hyper-parameter $r$ selected from 16 to 48 with step length 8; the $\lambda$ selected from 0.2 to 0.5 with step length 0.1; the number of extracted user/item historical contents $k$ is is set as $[3 \rightarrow 5 \rightarrow 7 \rightarrow 10]$.
%
%
%
%
%
%
Among all methods, we use the Adam~\cite{adam} to update parameters.

\subsection{Performance Comparisons}

Table \ref{main_exp} shows the performance of LoID on eleven datasets. 
%
All the experiments are repeated
5 times to sample historical contents in Section \ref{historical}, and we report the (average) test MSE obtained when the validation MSE is the lowest.
We observe that LoID achieves significant improvement over all SOTA baseline methods.
We note that aspect-aware methods such as ALFM and ANR consistently outperform DeepCoNN and D-Attn. 
We attribute this to the fact that DeepCoNN and D-Attn lack a comprehensive model for the intricate decision-making process in user-item interactions.
In addition, RGCL outperforms single-aspect and multi-aspect methods, indicating superior performance of graph neural networks. 
However, BiGI, though GNNs-based methods, falls short compared to single-aspect methods, without making use of reviews. 

We observe that the results become worse when removing the contrastive learning module. 
%
%
Moreover, LoID (Elec) results are superior to LoID.
This demonstrates the effectiveness of our LoRA-based plugin idea.  
%
When multiple LoRAs are employed, the results outperform when using single LoRA. 
%
We believe more LoRAs, more users' behavior and preference data.

\subsection{Discussion of Domain Correlation Effect}
We investigate whether the performance is correlated to the similarity between the target and source domains. 
We select $n$ reviews from each dataset randomly, employ the Sentence-BERT~\cite{sbert}, to quantify the cosine similarity between the domains.
In our experiments, we set $n$ to be 100. 
%
%
Then, we show the improvements compared with the original LoID (as shown in Table \ref{comparison_datasets}).
We conclude that as the similarity between the source and target domains increases, there is a corresponding rise in the model's performance.
%



\subsection{Discussion of Different LLMs}
\begin{table}[t]
    \centering
    \footnotesize
    \caption{Performance comparison of Different LLMs.}
    \label{LLM}
    \setlength\tabcolsep{3.3pt}{
        \begin{tabular}{lccc|ccc}
        \toprule
        \multirow{2}{*}{Datasets} & \multicolumn{3}{c|}{BERT} &\multicolumn{3}{c}{GPT2-medium}\\ \cline{2-7}&Precision &Recall &F1-score &Precision &Recall &F1-score \\
        \hline
        Amazon Instant Video & 0.7402 & 0.7578 & 0.7472 & 0.615 & 0.5999 & 0.606\\
        Digital Music &0.7692 & 0.7767 & 0.7712 & 0.7244 & 0.7136 & 0.7183\\
        Musical Instruments & 0.7376 & 0.7436 & 0.7382 & 0.5534 & 0.5227 & 0.5332\\
        \bottomrule
        \end{tabular}
    }
\vspace{-0.5cm}
\end{table}

We investigate the impact of different LLMs in the pre-training phase.
%
We consider two distinct paradigms, i.e., GPT and BERT. 
Due to prompt-based GPT being unable to predict floating-point numbers, thus we formulate the prompt:
{\footnotesize\textbf{Input template:} Give some example: \{\textit{content1}\} is score \{\textit{score1}\}, \{\textit{content2}\} is score \{\textit{score2}\}. Guess the score (The score should be between 1 and 5, where 1 means the lowest score, and 5 means the highest score) of \{\textit{current content}\}, we think the score is? $\quad$ \textbf{Target template:} \{\textit{score}\}, \{\textit{explanation}\}}.
To ensure a fair comparison, we tuned them by LoRA and adopted PRF (Precision, Recall, and F1-score) as the evaluation protocol.
Table \ref{LLM} presents the results, indicating that the BERT outperforms GPT2-medium. 
%

\section{Conclusions}
This paper introduces a simple yet effective approach named LoID, which includes two major components.
(1) `Pre-training plugin', we propose a flexible plugin framework that could transfer different domain semantic knowledge without re-training.
(2) `Aligning semantic/ID space', we devise a novel attention mechanism to connect the semantic and ID space, making our model easily applied in industrial RS.
Extensive experiments reveal that LoID surpasses existing SOTA methods, and in-depth analyses underscore the effectiveness of our model components.
In the future, we will explore the vision signal to improve our model ability.

\section*{Acknowledgement}
This work is supported by the National Natural Science Foundation of China under Grant No. 62206128 and the Undergraduate Research Training Program of Nanjing University
of Science and Technology (established in 2023) under Grant No. 2023066021B.

\balance
\bibliographystyle{ACM-Reference-Format}
\bibliography{sample-base-extend.bib}


\begin{thebibliography}{27}


\ifx \showCODEN    \undefined \def \showCODEN     #1{\unskip}     \fi
\ifx \showDOI      \undefined \def \showDOI       #1{#1}\fi
\ifx \showISBNx    \undefined \def \showISBNx     #1{\unskip}     \fi
\ifx \showISBNxiii \undefined \def \showISBNxiii  #1{\unskip}     \fi
\ifx \showISSN     \undefined \def \showISSN      #1{\unskip}     \fi
\ifx \showLCCN     \undefined \def \showLCCN      #1{\unskip}     \fi
\ifx \shownote     \undefined \def \shownote      #1{#1}          \fi
\ifx \showarticletitle \undefined \def \showarticletitle #1{#1}   \fi
\ifx \showURL      \undefined \def \showURL       {\relax}        \fi
\providecommand\bibfield[2]{#2}
\providecommand\bibinfo[2]{#2}
\providecommand\natexlab[1]{#1}
\providecommand\showeprint[2][]{arXiv:#2}

\bibitem[Brown et~al\mbox{.}(2020)]%
        {gpt-3}
\bibfield{author}{\bibinfo{person}{Tom Brown}, \bibinfo{person}{Benjamin Mann}, \bibinfo{person}{Nick Ryder}, \bibinfo{person}{Melanie Subbiah}, \bibinfo{person}{Jared~D Kaplan}, \bibinfo{person}{Prafulla Dhariwal}, \bibinfo{person}{Arvind Neelakantan}, \bibinfo{person}{Pranav Shyam}, \bibinfo{person}{Girish Sastry}, \bibinfo{person}{Amanda Askell}, {et~al\mbox{.}}} \bibinfo{year}{2020}\natexlab{}.
\newblock \showarticletitle{Language models are few-shot learners}.
\newblock \bibinfo{journal}{\emph{Advances In Neural Information Processing Systems}} (\bibinfo{year}{2020}).
\newblock


\bibitem[Cao et~al\mbox{.}(2021)]%
        {BiGI}
\bibfield{author}{\bibinfo{person}{Jiangxia Cao}, \bibinfo{person}{Xixun Lin}, \bibinfo{person}{Shu Guo}, \bibinfo{person}{Luchen Liu}, \bibinfo{person}{Tingwen Liu}, {and} \bibinfo{person}{Bin Wang}.} \bibinfo{year}{2021}\natexlab{}.
\newblock \showarticletitle{Bipartite graph embedding via mutual information maximization}. In \bibinfo{booktitle}{\emph{ACM International Conference on Web Search and Data Mining (WSDM)}}.
\newblock


\bibitem[Catherine and Cohen(2017)]%
        {transnets}
\bibfield{author}{\bibinfo{person}{Rose Catherine} {and} \bibinfo{person}{William Cohen}.} \bibinfo{year}{2017}\natexlab{}.
\newblock \showarticletitle{Transnets: Learning to transform for recommendation}. In \bibinfo{booktitle}{\emph{ACM Conference on Recommender Systems (RecSys)}}.
\newblock


\bibitem[Chen et~al\mbox{.}(2018)]%
        {regression}
\bibfield{author}{\bibinfo{person}{Chong Chen}, \bibinfo{person}{Min Zhang}, \bibinfo{person}{Yiqun Liu}, {and} \bibinfo{person}{Shaoping Ma}.} \bibinfo{year}{2018}\natexlab{}.
\newblock \showarticletitle{Neural attentional rating regression with review-level explanations}. In \bibinfo{booktitle}{\emph{The World Wide Web Conference (WWW)}}.
\newblock


\bibitem[Cheng et~al\mbox{.}(2016)]%
        {widedeep}
\bibfield{author}{\bibinfo{person}{Heng-Tze Cheng}, \bibinfo{person}{Levent Koc}, \bibinfo{person}{Jeremiah Harmsen}, \bibinfo{person}{Tal Shaked}, \bibinfo{person}{Tushar Chandra}, \bibinfo{person}{Hrishi Aradhye}, \bibinfo{person}{Glen Anderson}, \bibinfo{person}{Greg Corrado}, \bibinfo{person}{Wei Chai}, \bibinfo{person}{Mustafa Ispir}, {et~al\mbox{.}}} \bibinfo{year}{2016}\natexlab{}.
\newblock \showarticletitle{Wide \& deep learning for recommender systems}. In \bibinfo{booktitle}{\emph{ACM Conference on Recommender Systems Workshop}}.
\newblock


\bibitem[Cheng et~al\mbox{.}(2018)]%
        {alfm}
\bibfield{author}{\bibinfo{person}{Zhiyong Cheng}, \bibinfo{person}{Ying Ding}, \bibinfo{person}{Lei Zhu}, {and} \bibinfo{person}{Mohan Kankanhalli}.} \bibinfo{year}{2018}\natexlab{}.
\newblock \showarticletitle{Aspect-aware latent factor model: Rating prediction with ratings and reviews}. In \bibinfo{booktitle}{\emph{The World Wide Web Conference (WWW)}}.
\newblock


\bibitem[Chin et~al\mbox{.}(2018)]%
        {anr}
\bibfield{author}{\bibinfo{person}{Jin~Yao Chin}, \bibinfo{person}{Kaiqi Zhao}, \bibinfo{person}{Shafiq Joty}, {and} \bibinfo{person}{Gao Cong}.} \bibinfo{year}{2018}\natexlab{}.
\newblock \showarticletitle{ANR: Aspect-based neural recommender}. In \bibinfo{booktitle}{\emph{ACM International Conference on Information and Knowledge Management (CIKM)}}.
\newblock


\bibitem[Covington et~al\mbox{.}(2016)]%
        {youtubenet}
\bibfield{author}{\bibinfo{person}{Paul Covington}, \bibinfo{person}{Jay Adams}, {and} \bibinfo{person}{Emre Sargin}.} \bibinfo{year}{2016}\natexlab{}.
\newblock \showarticletitle{Deep neural networks for youtube recommendations}. In \bibinfo{booktitle}{\emph{ACM Conference on Recommender Systems (RecSys)}}.
\newblock


\bibitem[Devlin et~al\mbox{.}(2018)]%
        {bert}
\bibfield{author}{\bibinfo{person}{Jacob Devlin}, \bibinfo{person}{Ming-Wei Chang}, \bibinfo{person}{Kenton Lee}, {and} \bibinfo{person}{Kristina Toutanova}.} \bibinfo{year}{2018}\natexlab{}.
\newblock \showarticletitle{Bert: Pre-training of deep bidirectional transformers for language understanding}.
\newblock \bibinfo{journal}{\emph{ArXiv}} (\bibinfo{year}{2018}).
\newblock


\bibitem[Fu et~al\mbox{.}(2024)]%
        {adapter}
\bibfield{author}{\bibinfo{person}{Junchen Fu}, \bibinfo{person}{Fajie Yuan}, \bibinfo{person}{Yu Song}, \bibinfo{person}{Zheng Yuan}, \bibinfo{person}{Mingyue Cheng}, \bibinfo{person}{Shenghui Cheng}, \bibinfo{person}{Jiaqi Zhang}, \bibinfo{person}{Jie Wang}, {and} \bibinfo{person}{Yunzhu Pan}.} \bibinfo{year}{2024}\natexlab{}.
\newblock \showarticletitle{Exploring adapter-based transfer learning for recommender systems: Empirical studies and practical insights}. In \bibinfo{booktitle}{\emph{ACM International Conference on Web Search and Data Mining (WSDM)}}.
\newblock


\bibitem[Hu et~al\mbox{.}(2021)]%
        {lora}
\bibfield{author}{\bibinfo{person}{Edward~J Hu}, \bibinfo{person}{Yelong Shen}, \bibinfo{person}{Phillip Wallis}, \bibinfo{person}{Zeyuan Allen-Zhu}, \bibinfo{person}{Yuanzhi Li}, \bibinfo{person}{Shean Wang}, \bibinfo{person}{Lu Wang}, {and} \bibinfo{person}{Weizhu Chen}.} \bibinfo{year}{2021}\natexlab{}.
\newblock \showarticletitle{Lora: Low-rank adaptation of large language models}.
\newblock \bibinfo{journal}{\emph{ArXiv}} (\bibinfo{year}{2021}).
\newblock


\bibitem[Kingma and Ba(2015)]%
        {adam}
\bibfield{author}{\bibinfo{person}{P.~Diederik Kingma} {and} \bibinfo{person}{Lei~Jimmy Ba}.} \bibinfo{year}{2015}\natexlab{}.
\newblock \showarticletitle{Adam: A Method for Stochastic Optimization}. In \bibinfo{booktitle}{\emph{International Conference on Learning Representations (ICLR)}}.
\newblock


\bibitem[Li et~al\mbox{.}(2023)]%
        {qformer}
\bibfield{author}{\bibinfo{person}{Junnan Li}, \bibinfo{person}{Dongxu Li}, \bibinfo{person}{Silvio Savarese}, {and} \bibinfo{person}{Steven Hoi}.} \bibinfo{year}{2023}\natexlab{}.
\newblock \showarticletitle{Blip-2: Bootstrapping language-image pre-training with frozen image encoders and large language models}.
\newblock \bibinfo{journal}{\emph{ArXiv}} (\bibinfo{year}{2023}).
\newblock


\bibitem[Liangwei~Yang(2023)]%
        {DGRec}
\bibfield{author}{\bibinfo{person}{Yunzhe Tao Jiankai Sun Xiaolong Liu Philip S. Yu Taiqing~Wang Liangwei~Yang, Shengjie~Wang}.} \bibinfo{year}{2023}\natexlab{}.
\newblock \showarticletitle{DGRec: Graph Neural Network for Recommendation with Diversified Embedding Generation}. In \bibinfo{booktitle}{\emph{ACM International Conference on Web Search and Data Mining (WSDM)}}.
\newblock


\bibitem[Pi et~al\mbox{.}(2020)]%
        {sim}
\bibfield{author}{\bibinfo{person}{Qi Pi}, \bibinfo{person}{Guorui Zhou}, \bibinfo{person}{Yujing Zhang}, \bibinfo{person}{Zhe Wang}, \bibinfo{person}{Lejian Ren}, \bibinfo{person}{Ying Fan}, \bibinfo{person}{Xiaoqiang Zhu}, {and} \bibinfo{person}{Kun Gai}.} \bibinfo{year}{2020}\natexlab{}.
\newblock \showarticletitle{Search-based user interest modeling with lifelong sequential behavior data for click-through rate prediction}. In \bibinfo{booktitle}{\emph{ACM International Conference on Information and Knowledge Management (CIKM)}}.
\newblock


\bibitem[Radford et~al\mbox{.}(2021)]%
        {clip}
\bibfield{author}{\bibinfo{person}{Alec Radford}, \bibinfo{person}{Jong~Wook Kim}, \bibinfo{person}{Chris Hallacy}, \bibinfo{person}{Aditya Ramesh}, \bibinfo{person}{Gabriel Goh}, \bibinfo{person}{Sandhini Agarwal}, \bibinfo{person}{Girish Sastry}, \bibinfo{person}{Amanda Askell}, \bibinfo{person}{Pamela Mishkin}, \bibinfo{person}{Jack Clark}, {et~al\mbox{.}}} \bibinfo{year}{2021}\natexlab{}.
\newblock \showarticletitle{Learning transferable visual models from natural language supervision}. In \bibinfo{booktitle}{\emph{International Conference on Machine Learning (ICML)}}.
\newblock


\bibitem[Reimers and Gurevych(2019)]%
        {sbert}
\bibfield{author}{\bibinfo{person}{Nils Reimers} {and} \bibinfo{person}{Iryna Gurevych}.} \bibinfo{year}{2019}\natexlab{}.
\newblock \showarticletitle{Sentence-bert: Sentence embeddings using siamese bert-networks}.
\newblock \bibinfo{journal}{\emph{ArXiv}} (\bibinfo{year}{2019}).
\newblock


\bibitem[Rendle(2010)]%
        {fm}
\bibfield{author}{\bibinfo{person}{Steffen Rendle}.} \bibinfo{year}{2010}\natexlab{}.
\newblock \showarticletitle{Factorization machines}. In \bibinfo{booktitle}{\emph{IEEE International Conference on Data Mining}}.
\newblock


\bibitem[Seo et~al\mbox{.}(2017)]%
        {dattn}
\bibfield{author}{\bibinfo{person}{Sungyong Seo}, \bibinfo{person}{Jing Huang}, \bibinfo{person}{Hao Yang}, {and} \bibinfo{person}{Yan Liu}.} \bibinfo{year}{2017}\natexlab{}.
\newblock \showarticletitle{Interpretable convolutional neural networks with dual local and global attention for review rating prediction}. In \bibinfo{booktitle}{\emph{ACM Conference on Recommender Systems (RecSys)}}.
\newblock


\bibitem[Shuai et~al\mbox{.}(2022)]%
        {RGCL}
\bibfield{author}{\bibinfo{person}{Jie Shuai}, \bibinfo{person}{Kun Zhang}, \bibinfo{person}{Le Wu}, \bibinfo{person}{Peijie Sun}, \bibinfo{person}{Richang Hong}, \bibinfo{person}{Meng Wang}, {and} \bibinfo{person}{Yong Li}.} \bibinfo{year}{2022}\natexlab{}.
\newblock \showarticletitle{A review-aware graph contrastive learning framework for recommendation}. In \bibinfo{booktitle}{\emph{ACM International Conference on Research on Development in Information Retrieval (SIGIR)}}.
\newblock


\bibitem[Smith and Linden(2017)]%
        {amazon}
\bibfield{author}{\bibinfo{person}{Brent Smith} {and} \bibinfo{person}{Greg Linden}.} \bibinfo{year}{2017}\natexlab{}.
\newblock \showarticletitle{Two decades of recommender systems at Amazon. com}.
\newblock \bibinfo{journal}{\emph{IEEE International Computing}} (\bibinfo{year}{2017}).
\newblock


\bibitem[Vaswani et~al\mbox{.}(2017)]%
        {attention}
\bibfield{author}{\bibinfo{person}{Ashish Vaswani}, \bibinfo{person}{Noam Shazeer}, \bibinfo{person}{Niki Parmar}, \bibinfo{person}{Jakob Uszkoreit}, \bibinfo{person}{Llion Jones}, \bibinfo{person}{Aidan~N Gomez}, \bibinfo{person}{{\L}ukasz Kaiser}, {and} \bibinfo{person}{Illia Polosukhin}.} \bibinfo{year}{2017}\natexlab{}.
\newblock \showarticletitle{Attention is all you need}.
\newblock \bibinfo{journal}{\emph{Advances In Neural Information Processing Systems}} (\bibinfo{year}{2017}).
\newblock


\bibitem[Ying et~al\mbox{.}(2018)]%
        {pinsage}
\bibfield{author}{\bibinfo{person}{Rex Ying}, \bibinfo{person}{Ruining He}, \bibinfo{person}{Kaifeng Chen}, \bibinfo{person}{Pong Eksombatchai}, \bibinfo{person}{William~L Hamilton}, {and} \bibinfo{person}{Jure Leskovec}.} \bibinfo{year}{2018}\natexlab{}.
\newblock \showarticletitle{Graph Convolutional Neural Networks for Web-Scale Recommender Systems}. In \bibinfo{booktitle}{\emph{ACM Knowledge Discovery and Data Mining (KDD)}}.
\newblock


\bibitem[Yu et~al\mbox{.}(2023)]%
        {DARE}
\bibfield{author}{\bibinfo{person}{Le Yu}, \bibinfo{person}{Bowen Yu}, \bibinfo{person}{Haiyang Yu}, \bibinfo{person}{Fei Huang}, {and} \bibinfo{person}{Yongbin Li}.} \bibinfo{year}{2023}\natexlab{}.
\newblock \showarticletitle{Language models are super mario: Absorbing abilities from homologous models as a free lunch}.
\newblock \bibinfo{journal}{\emph{ArXiv}} (\bibinfo{year}{2023}).
\newblock


\bibitem[Yuan et~al\mbox{.}(2023)]%
        {id}
\bibfield{author}{\bibinfo{person}{Zheng Yuan}, \bibinfo{person}{Fajie Yuan}, \bibinfo{person}{Yu Song}, \bibinfo{person}{Youhua Li}, \bibinfo{person}{Junchen Fu}, \bibinfo{person}{Fei Yang}, \bibinfo{person}{Yunzhu Pan}, {and} \bibinfo{person}{Yongxin Ni}.} \bibinfo{year}{2023}\natexlab{}.
\newblock \showarticletitle{Where to go next for recommender systems? id-vs. modality-based recommender models revisited}. In \bibinfo{booktitle}{\emph{ACM International Conference on Research on Development in Information Retrieval (SIGIR)}}.
\newblock


\bibitem[Zheng et~al\mbox{.}(2017)]%
        {deepconn}
\bibfield{author}{\bibinfo{person}{Lei Zheng}, \bibinfo{person}{Vahid Noroozi}, {and} \bibinfo{person}{Philip~S Yu}.} \bibinfo{year}{2017}\natexlab{}.
\newblock \showarticletitle{Joint deep modeling of users and items using reviews for recommendation}. In \bibinfo{booktitle}{\emph{ACM International Conference on Web Search and Data Mining (WSDM)}}.
\newblock


\bibitem[Zhou et~al\mbox{.}(2018)]%
        {din}
\bibfield{author}{\bibinfo{person}{Guorui Zhou}, \bibinfo{person}{Xiaoqiang Zhu}, \bibinfo{person}{Chenru Song}, \bibinfo{person}{Ying Fan}, \bibinfo{person}{Han Zhu}, \bibinfo{person}{Xiao Ma}, \bibinfo{person}{Yanghui Yan}, \bibinfo{person}{Junqi Jin}, \bibinfo{person}{Han Li}, {and} \bibinfo{person}{Kun Gai}.} \bibinfo{year}{2018}\natexlab{}.
\newblock \showarticletitle{Deep interest network for click-through rate prediction}. In \bibinfo{booktitle}{\emph{ACM Knowledge Discovery and Data Mining (KDD)}}.
\newblock


\end{thebibliography}
\end{document}